\documentclass[reprint,amsmath,amssymb,aps,showpacs,pra]{revtex4-1}
\usepackage{graphicx}% Include figure files
\usepackage{dcolumn}% Align table columns on decimal point
\usepackage{bm}% bold math
\usepackage[usenames,dvipsnames]{color}
\usepackage[utf8]{inputenc}
\usepackage{hyperref}% add hypertext capabilities
\graphicspath{{./Files/PDFs/}}

\begin{document}

%=========================================================================================

\title{Coherent nonlinear oscillations in magnetohydrodynamic plasma}

%\vspace{8mm}
%\large
\author{Rupak Mukherjee} 
\email{rupak@ipr.res.in; rupakmukherjee01@gmail.com}
\author {Rajaraman Ganesh}
%\email{ganesh@ipr.res.in}
\author{Abhijit Sen} 
\email{abhijit@ipr.res.in}
\affiliation{Institute for Plasma Research, HBNI, Bhat, Gandhinagar - 382428, India}

\begin{abstract}
Single fluid magnetohydrodynamic (MHD) equations have been studied through direct numerical simulations (DNS) using pseudo-spectral methods in two as well as three spatial dimensions. At Alfv\'en resonance, a reversible periodic exchange of energy between kinetic and magnetic variables is observed. The oscillations are identified as nonlinear dispersionless Alfv\'en waves that have been predicted earlier on theoretical grounds but not observed in large scale numerical simulations. A systematic study of their occurrence for various initial conditions and a range of Alfven velocities is carried out. An analysis based on a finite mode representation of the incompressible single fluid MHD equations in two spatial dimensions reproduces the essential features of these oscillations.
\end{abstract}

\maketitle

%=========================================================================================

\section{Introduction}

Understanding kinetic energy cascade for different initial spectra of perturbations has been a longstanding challenge in the field of fluid dynamics. Of late several direct numerical simulation (DNS) techniques have provided powerful tools to study the time evolution of the kinetic energy spectra of model fluids governed by the Navier-Stokes equation. The fluid equations represented as an infinite dimensional system are hard to solve analytically \cite{morrison:2006}. The complexity increases manifold when the fluid is charged and hence can interact with external as well as internal electric and magnetic fields. The self-generated magnetic field dynamically affects the fluid flow through the Lorentz force term in the momentum equation. Thus the energy accumulated in any mode may be expected to find new routes via nonlinear cascades to other modes through transformation of kinetic to magnetic energy and vice versa.\\

In the presence of weak resistivity, the MHD model is known to support irreversible conversion of magnetic energy into fluid kinetic energy through the process of magnetic reconnection as well as conversion of kinetic energy into mean large scale magnetic field through a dynamo mechanism. Therefore it is interesting to ask oneself, for a given fluid type and magnetic field strength, whether there are fluid flow profiles which do neither - that is, neither does the flow generate a mean magnetic field nor does the magnetic field energy get converted to flow energy. Can there be a nearly ``reversible" process in the form coherent nonlinear oscillations that cycle the energy back and forth between the magnetic and kinetic form. In this work we report the existence of such large amplitude oscillations for a wide range of initial flow speeds or Alfven Mach numbers. The oscillations are identified as nonlinear dispersionless Alfven waves and are found to be excited only for special initial conditions consisting of sinusoidal wave-forms in fluid flows on a homogeneous ambient magnetic field.\\

The existence of such dispersionless nonlinear oscillations are known for some time \cite{alfven:1963} which have been revisited more recently by Yoshida \cite{yoshida:2012} and Abdelhamid \& Yoshida \cite{abdelhamid:2016} from a point of view of Hamiltonian-Casimir formulation of ideal MHD and its extended models. They had also suggested the necessity of special initial wave forms for sustaining such oscillations. However, to the best of our knowledge, there have been no systematic study of observations of such oscillations in direct numerical simulations of the model equations and our present work provides one.\\

Our detailed direct numerical simulation (DNS) studies are carried out for both three dimensional as well as two dimensional single fluid MHD equations using an in-house developed MPI parallel pseudo-spectral code  called MHD3D. We observe the periodic exchange of kinetic and magnetic energy components at the Alfven frequency for different sets of divergenceless initial flow profiles such as the  the Orszag-Tang flow in two dimensions and the Arnold-Beltrami-Childress flow in three dimensions. The oscillations are seen to be excited only for initial sinusoidal wave-forms in fluid flows on a homogeneous ambient magnetic field. Other initial conditions such as non-sinusoidal perturbations in fluid flows or sinusoidal perturbations in the magnetic component and a uniform fluid flow do not give rise to these coherent oscillations. Our findings are thus in agreement with the theoretical predictions of the ideal MHD model and provide a systematic evidence of the existence of the nonlinear dispersionless Alfven waves in a weakly non-ideal model. The small non-ideal terms of resistivity and viscosity provide a weak damping of these oscillations.\\

The shell averaged kinetic and magnetic energy spectra show that the energy is primarily confined in the large scales associated with the system. However, both the spectra at different times show that the energy keeps oscillating between kinetic and magnetic modes. Motivated from this numerical observation, a finite mode representation of the field variables is carried out for the incompressible MHD equations in two dimensions. The single fluid MHD equations when expanded in a Galerkin truncated mode fashion, give rise to a set of nonlinear coupled ordinary differential equations. The solution of the system of such equations is shown to capture the essential features of the nonlinear oscillations in support of the pseudo-spectral DNS results.\\

The paper is organized as follows. Section \ref{sec:eqn} presents the equations that are time evolved in the code MHD3D described in Section \ref{sec:code}. In Section \ref{sec:parameter} we provide all the parameters in which the code is run and Section \ref{sec:DNS} provides all the direct numerical simulation (DNS) results we have obtained in two and three dimensions. The analysis of the DNS data is done in Section \ref{sec:Analysis} and Section \ref{sec:galerkin} describes an analytical two dimensional finite mode representation of the DNS results. Section \ref{sec:summary} summarises all the results and analysis performed in this paper.

%=========================================================================================

\section{Governing Equations} 
\label{sec:eqn}
The basic equations governing the dynamics of the magnetohydrodynamic fluid are as follows: 
\begin{eqnarray}
&& \label{density} \frac{\partial \rho}{\partial t} + \vec{\nabla} \cdot \left(\rho \vec{u}\right) = 0\\
&& \frac{\partial (\rho \vec{u})}{\partial t} + \vec{\nabla} \cdot \left[ \rho \vec{u} \otimes \vec{u} + \left(P + \frac{B^2}{2}\right){\bf{I}} - \vec{B}\otimes\vec{B} \right]\nonumber \\
&& \label{velocity} ~~~~~~~~~ = \mu \nabla^2 \vec{u} + \rho\vec{f}\\
&& \label{Bfield} \frac{\partial \vec{B}}{\partial t} + \vec{\nabla} \cdot \left( \vec{u} \otimes \vec{B} - \vec{B} \otimes \vec{u}\right) = \eta \nabla^2 \vec{B}\\
&& \label{pressure} P = C_s^2 \rho
\end{eqnarray}
In this system of equations, $\rho$, $\vec{u}$, $P$ and $\vec{B}$ are the density, velocity, kinetic pressure and magnetic field of a fluid element respectively. $\mu$ and $\eta$ denote the coefficient of kinematic viscosity and magnetic resistivity. We assume $\mu$ and $\eta$ are constants over space and time. The symbol ``$\otimes$'' represents the dyadic between two vector quantities.\\

The kinetic Reynolds number ($Re$) and magnetic Reynolds number ($Rm$) are defined as $Re = \frac{U_0 L}{\mu}$ and $Rm = \frac{U_0 L}{\eta}$ where $U_0$ is the maximum velocity of the fluid system to start with and $L$ is the system length.\\

We also define the sound speed of the fluid as $C_s = \frac{U_0}{M_s}$, where, $M_s$ is the sonic Mach number of the fluid. The Alfven speed is determined from $V_A = \frac{U_0}{M_A}$ where $M_A$ is the Alfven Mach number of the plasma. The initial magnetic field present in the plasma is determined from the relation $B_0 = V_A \sqrt{\rho_0}$, where, $\rho_0$ is the initial density profile of the fluid.\\

The length is normalised to the system size or box length ($L$) considered in the simulation and velocity is normalised to sound speed of the medium $\left(C_s = \frac{U_0}{M_s}\right)$.
%=========================================================================================

\section{Details of MHD3D code}
\label{sec:code}
In order to simulate the plasma dynamics governed through the MHD equations (Equation \ref{density}-\ref{pressure}), a DNS code, MHD3D, has been developed in-house at Institute for Plasma Research. MHD3D is an MPI parallel three dimensional weakly compressible, viscous, resistive magnetohydrodynamic code using pseudo-spectral technique to simulate a general scenario of three dimensional magnetohydrodynamic turbulence problem. The pseudo-spectral technique, one of the most accurate CFD techniques available till today, uses the FFTW libraries \cite{FFTW3:2005} which is one of the fastest Fourier Transform libraries developed recently. This technique is applied to calculate the spatial derivatives and to evaluate the non-linear terms involved in the basic underlying equations with a standard de-aliasing by $2/3$ truncation rule. The time derivative is solved using multiple time solvers viz. Adams-Bashforth, Runge-Kutta 4 and Predictor-Correcter algorithms and all the solvers have been checked to provide identical results.\\

For the two-dimensional hydrodynamic case, we reproduced the results obtained earlier by Drazin \cite{drazin:1961}, Ray \cite{ray:1982} and Keppens {\it et al} \cite{keppens:1999} for a Kelvin-Helmholtz like scenario in both the incompressible and compressible limits. The compressible magnetodydrodynamic solver in two dimensions is benchmarked with Keppens {\it et al} \cite{keppens:1999} and Orszag {\it et al} \cite{orszag:1979}. The three dimensional weakly compressible neutral fluid solver is benchmarked with the results of Samtaney {\it et al}\cite{ravi:2001} for the root mean square of velocity divergence and the skewness of velocity field for a decaying turbulence case. The three dimensional kinetic dynamo effect is matched with Galloway {\it et al} \cite{galloway:1986} in the case of ABC flow for a grid resolution of $64^3$. More benchmarking details and specifications about the code are given in earlier papers \cite{rupak:2018a,rupak:2018b,rupak:2018c}.

%=========================================================================================

\section{Parameter Details}
\label{sec:parameter}

We choose $L = 2 \pi$ periodic box as our simulation domain and fix $\rho_0 = 1$ and keep these parameters identical throughout our simulations. The initial magnitude of density ($\rho_0$) is known to affect the dynamics and growth rate of an instability in a compressible neutral fluid \cite{bayly:1992,terakado:2014}. However, in this report we keep the initial density fixed ($\rho_0 = 1$) for all the runs unless stated otherwise. 

For the two dimensional runs we choose a grid resolution of $N_x = N_y = 128$ and a time stepping interval $\delta t = 10^{-5}$. The kinetic and magnetic Reynolds numbers are chosen as $Re = Rm = 10^{-4}$. We run our code for the incompressible limit ($M_s = 0.01$). 

For three dimensional runs we choose a grid resolution of $N_x = N_y = Nz = 64$ and a time stepping interval $\delta t = 10^{-4}$. The kinetic and magnetic Reynolds numbers are chosen as $Re = Rm = 450$. For all cases we run our code for the incompressible limit ($M_s = 0.1$).

For both two and three dimensional runs we check our code with higher grid resolution ($N$) and smaller time stepping ($\delta t$) and find no significant variation in the results of our test runs.  

We start with incompressible solutions ($\vec{\nabla} \cdot \vec{u} = 0$) and time evolve the individual velocity profiles with very weak compressibility in the medium. The back coupling of the magnetic variables through Lorentz force term alters the regular well-known hydrodynamic evolution severely thereby giving rise to completely new dynamics in the system.

We choose $M_A = 1$ i.e. the mean velocity ($U_0$) is equal to the Alfven speed of the system ($V_A$) (i.e. Alfven resonance) for both the two as well as the three dimensional runs. 

For two dimensional flows the initial magnetic field profile is chosen as $B_x = B_y = B_0$ and that for three dimensional flows as $B_x = B_y = B_z = B_0$ where $B_0$ is a constant.

%=========================================================================================

\section{Simulation Results}
\label{sec:DNS}
%================================================

\subsection{Results for 2D Orszag-Tang Flow}

\begin{figure}
\includegraphics[scale=0.65]{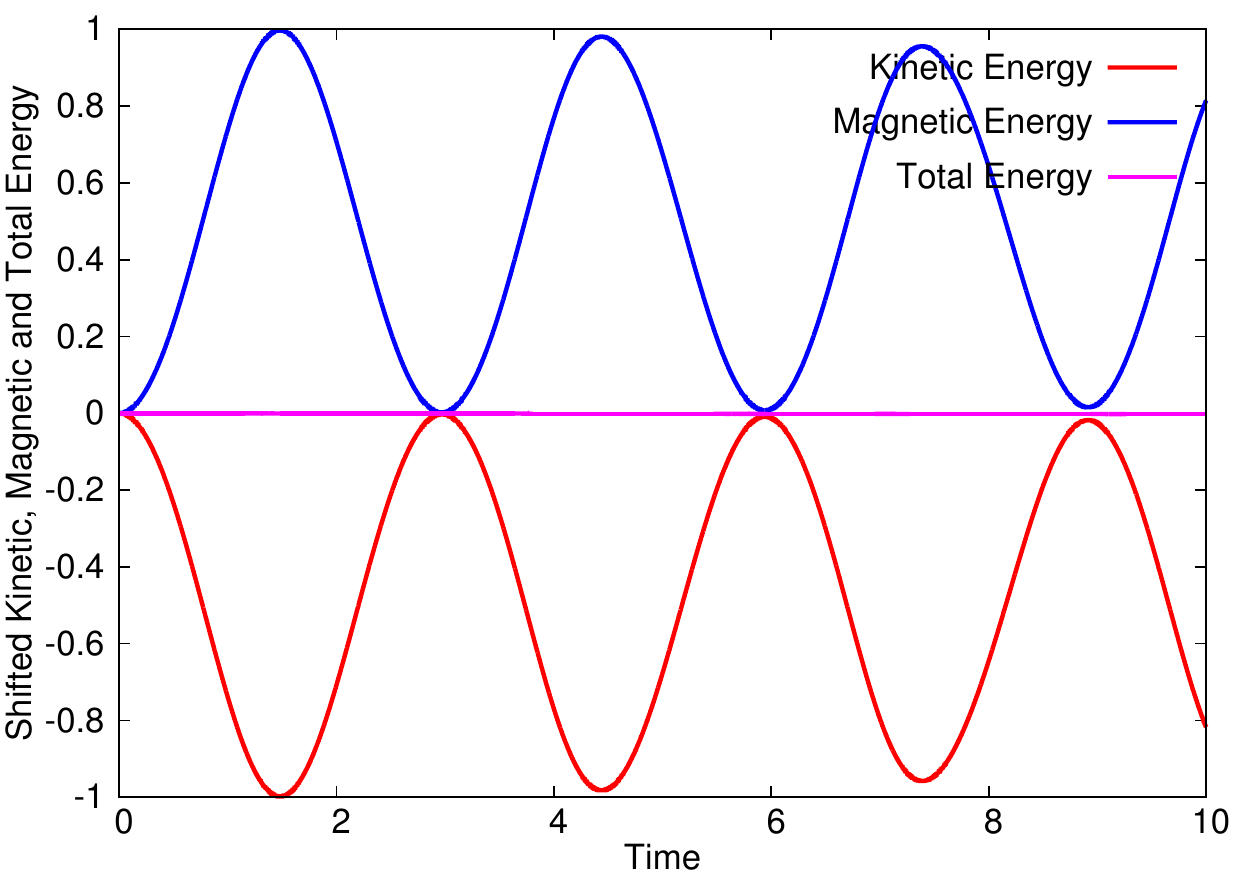}
\caption{Time evolution of shifted kinetic [$u^2(t) - u^2(0)$], magnetic [$B^2(t) - B^2(0)$] and total [$u^2(t) - u^2(0)$ + $B^2(t) - B^2(0)$] energy per grid for two dimensional Orszag-Tang flow with $N = 128^2$, $U_0 = 1$, $k_0 = 1$, $M_s = 0.01$, $M_A = 1$ and $Re = Rm = 10^{-4}$ in the absence of external forcing. The energy keeps on switching between kinetic and magnetic modes keeping the total energy conserved. It is found that the time period of oscillation is $T = 2.971$}
\label{energy_OT}
\end{figure}

For two dimensional Orszag-Tang flow the velocity profile is chosen as 
\begin{eqnarray}\label{OT2}
\begin{aligned}
& u_x = - A \sin(k_0 y)\\
& u_y = + A \sin(k_0 x)
\end{aligned}
\end{eqnarray} 
with $A = 1$ and $k_0 = 1$. The interchange of energy between kinetic and magnetic variables is plotted in Fig. \ref{energy_OT}. The x-axis represents the time evolution and the y-axis represents the shifted kinetic and magnetic energy where the shift indicates that the initial values of the corresponding variables are subtracted from the time evolution data. From Fig. \ref{energy_OT} it is observed that the time period of oscillation is $T = 2.971$.\\

A long time evolution of the system with several cycles of oscillations with $N = 256^2$, $\delta t = 10^{-6}$, $U_0 = 1$, $k_0 = 1$, $M_s = 0.01$, $M_A = 1$ and $Re = Rm = 10^{-5}$ is found to show numerical convergence. 

%================================================

\subsection{Results for 3D ABC Flow}

\begin{figure}
\includegraphics[scale=0.65]{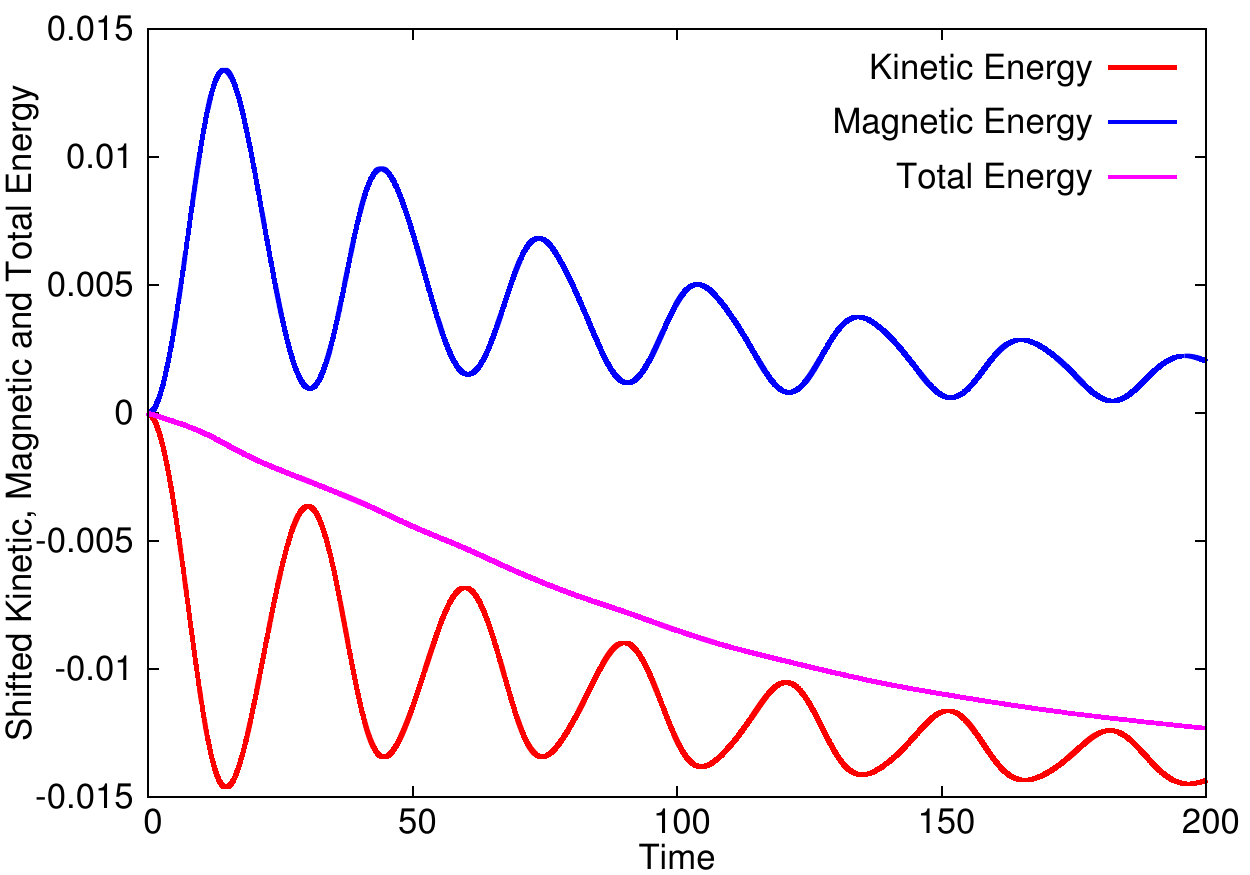}
\caption{Time evolution of shifted kinetic, magnetic and total energy per grid for three dimensional Arnold-Beltrami-Childress flow with $N = 64^3$, $U_0 = 0.1$, $A = B = C = 1$, $k_0 = 1$, $M_s = 0.1$, $M_A = 1$ and $Re = Rm = 450$ in the absence of external forcing. The energy keeps switching between kinetic and magnetic modes keeping the total energy conserved. The decay of total energy is solely due to viscous and resistive effects. The shift denotes the initial values of the kinetic and magnetic energies (at time $t = 0$) are subtracted from the time evolution data of the corresponding variables. The time period of oscillation is $T = 30.171$}
\label{energy_ABC}
\end{figure}

For three dimensional Arnold-Beltrami-Childress flow we choose the velocity profile as 
\begin{eqnarray}\label{ABC3}
\begin{aligned}
& u_x = U_0 [A \sin(k_0 z) + C \cos (k_0 y)]\\
& u_y = U_0 [B \sin(k_0 x) + A \cos (k_0 z)]\\
& u_z = U_0 [C \sin(k_0 y) + B \cos (k_0 x)]
\end{aligned}
\end{eqnarray} 
with $A = B = C = 1$. The interchange of energy between kinetic and magnetic variables is plotted in Fig. \ref{energy_ABC}. It is found that the time period of oscillation is $T = 30.171$.

%================================================================================================

\section{Analysis of DNS results}
\label{sec:Analysis}

The oscillations of the kinetic energy and magnetic energy are found to be related to the Alfven frequency of the system. In two dimensions, for $M_A = 1$ and $U_0 = 1$, the Alfven speed should be $V_A = \frac{U_0}{M_A} = 1$. The corresponding Alfven frequency should be $\omega_A = \frac{2\pi}{T_A} = \frac{2\pi V_A}{L} = 1$. Thus the corresponding energy should occur with a frequency twice the Alfven frequency (as the energy varies as square of velocity). So, the frequency of oscillation of kinetic and magnetic energy should be $\omega_{Th} = 2 \omega_A = 2$. From our numerical simulation we find that the frequency of oscillation is $\omega_{Num} = \frac{2\pi}{2.971} = 2.115$ which is close to the oscillation frequency due to Alfven waves ($\omega_{Th}$).\\

In three dimensions, for $M_A = 1$ and $U_0 = 0.1$, the Alfven speed should be $V_A = \frac{U_0}{M_A} = 0.1$. The corresponding Alfven frequency should be $\omega_A = \frac{2\pi}{T_A} = \frac{2\pi V_A}{L} = 0.1$. Thus, the frequency of oscillation of kinetic and magnetic energy should be $\omega_{Th} = 2 \omega_A = 0.2$. From our numerical simulation we find that, $\omega_{Num} = \frac{2\pi}{30.171} = 0.2083$ which is close to ($\omega_{Th}$).\\

However, the amplitudes of oscillation in all the cases are found to be very high, unlike the linear Alfven waves. For example, in three dimensional cases, the initial kinetic and magnetic energies are $\frac{1}{2} U_0^2 = 0.015 = \frac{1}{2} B_0^2$, for $M_A = \frac{u_0}{V_A} = 1$, $U_0 = 0.1$ and $B_0 = V_A \sqrt{\rho_0} = V_A \cdot 1 = 0.1$. The amplitude of oscillation of shifted kinetic and magnetic energy is around $0.014$. Thus the percentage change in energy is around $93\%$ which is very high indicating the strong non-linearity of the system. It is also found that, in the course of oscillation, almost all the energy is converted from one form to the other and is reproduced back.\\

%===========================

\subsection{Parameter Dependency of 3D results}

%===========================

\subsubsection{Effect of Alfven Speed ($M_A$)}

We know the time period oscillation of Alfven wave ($T_A$) varies linearly with $B_0$ which is directly proportional to $V_A = \frac{U_0}{M_A}$. Thus if, $M_A$ is increased, $T_A$ should decrease linearly. We analyse the frequency of oscillation for a series of $M_A$ values and find good agreement with this linearity prediction [Table \ref{M_A_vs_omega}].\\

\begin{figure}
\includegraphics[scale=0.65]{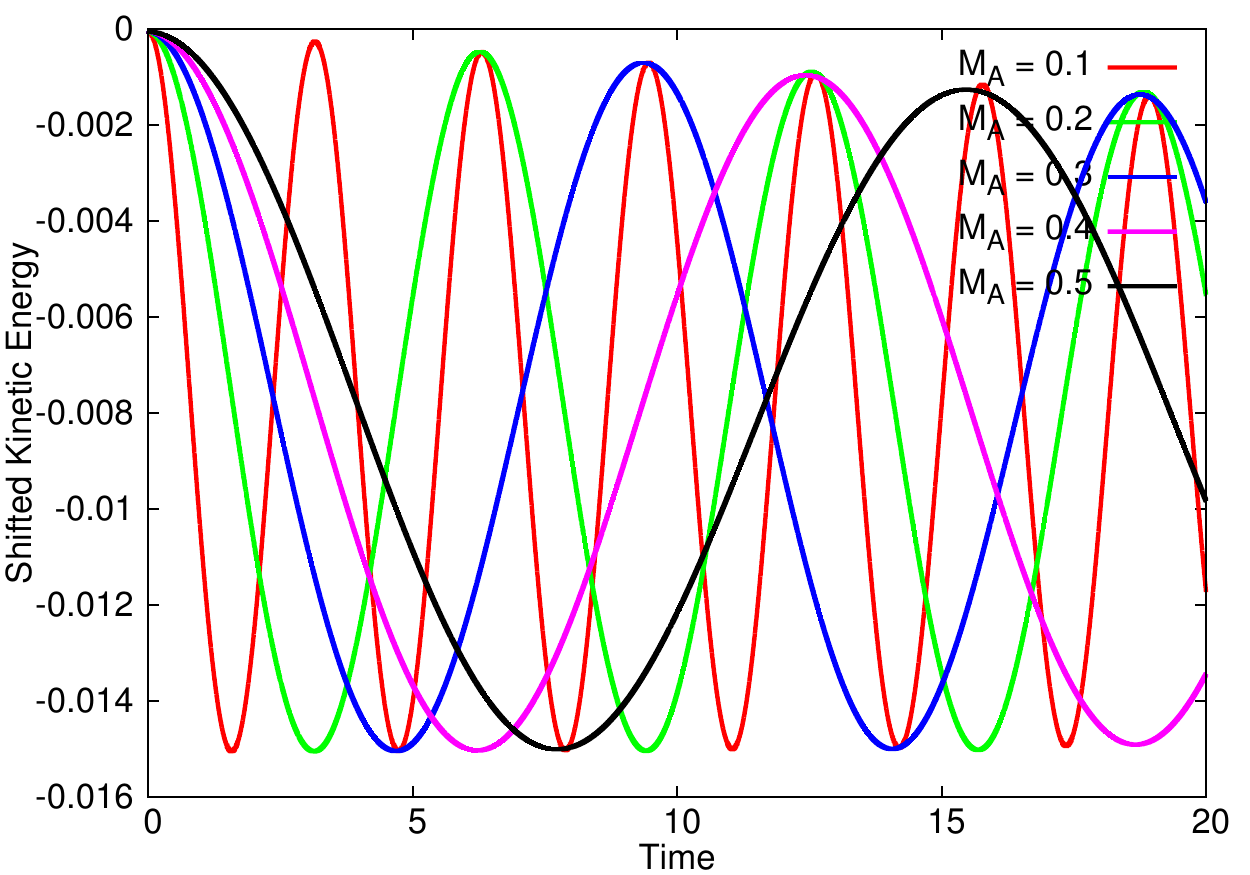}
\caption{Time evolution of shifted kinetic energy per grid for three dimensional Arnold-Beltrami-Childress flow with $N = 64^3$, $U_0 = 0.1$, $A = B = C = 1$, $k_0 = 1$, $M_s = 0.1$, $M_A = 0.1$ (red), $0.2$ (greed), $0.3$ (blue), $0.4$ (magenta), $0.5$ (black) and $Re = Rm = 450$ in the absence of external forcing. The frequency of oscillation decreases with the increase of $M_A$. The shift denotes the initial values of the kinetic and magnetic energies (at time $t = 0$) are subtracted from the time evolution data of the corresponding variables.}
\label{MA_low}
\end{figure}

\begin{figure}
\includegraphics[scale=0.65]{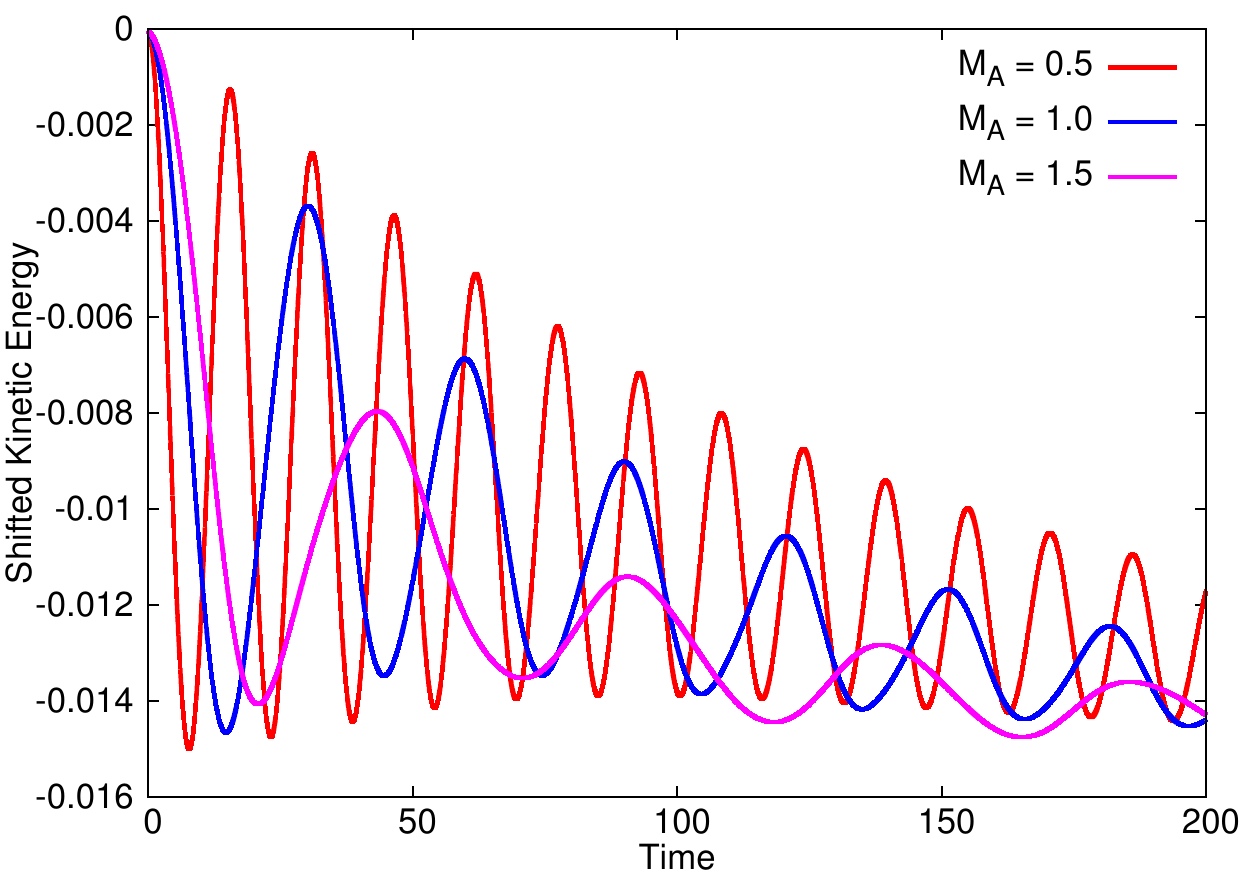}
\caption{Time evolution of shifted kinetic energy per grid for three dimensional Arnold-Beltrami-Childress flow with $N = 64^3$, $U_0 = 0.1$, $A = B = C = 1$, $k_0 = 1$, $M_s = 0.1$, $M_A = 0.5$ (red), $1$ (blue), $1.5$ (magenta) and $Re = Rm = 450$ in the absence of external forcing. The frequency of oscillation decreases with the increase of $M_A$. The shift denotes the initial values of the kinetic and magnetic energies (at time $t = 0$) are subtracted from the time evolution data of the corresponding variables.}
\label{MA}
\end{figure}

\begin{figure}
\includegraphics[scale=0.65]{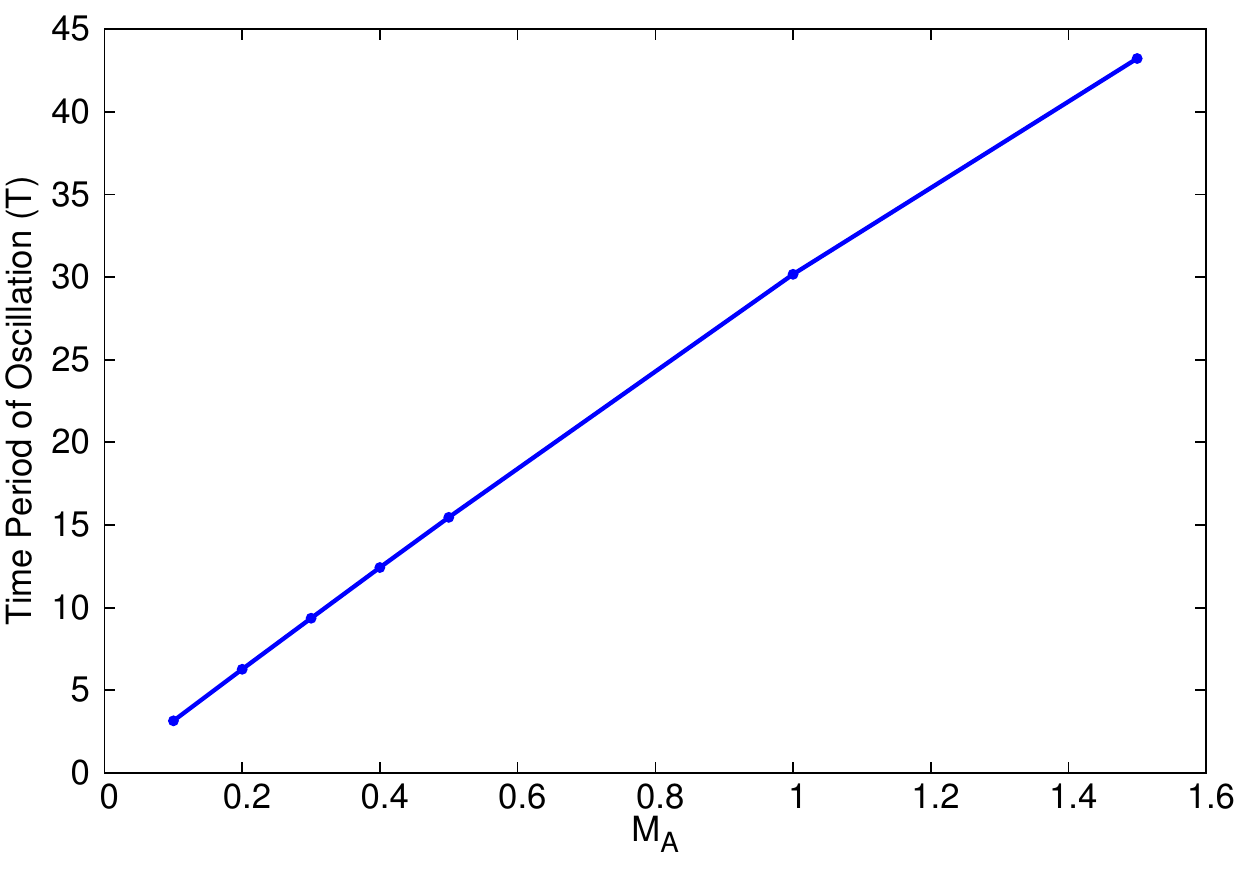}
\caption{Variation of time period of oscillation ($T$) of kinetic and magnetic energy with $M_A$. [see Table \ref{M_A_vs_omega}]}
\label{freq_MA}
\end{figure}

Below we provide some examples of the energy exchange process by changing the Alfven Mach number $M_A$ for $k_f = 1$. Fig.(\ref{MA_low}) represents the oscillation of kinetic energy for $M_A = 0.1, 0.2, 0.3, 0.4, 0.5$ and Fig.(\ref{MA}) represents the oscillation of kinetic energy for $M_A = 0.5,1.0, 1.5$. From Fig. \ref{freq_MA} it is evident that the frequency of oscillation decreases linearly with the increase of $M_A$.\\

\begin{table}[h!] 
\centering
\begin{tabular}{ |c|c| }
 \hline
 ~~~~$M_A$~~~~ & Time Period of Oscillation (T) \\
 \hline
 ~0.01 ~&~ 0.321~\\
 ~0.02 ~&~ 0.699~\\
 ~0.03 ~&~ 0.974~\\
 ~0.04 ~&~ 1.295~\\
 ~0.05 ~&~ 3.932~\\
 ~0.06 ~&~ 1.914~\\
 ~0.07 ~&~ 2.221~\\
 ~0.08 ~&~ 2.479~\\
 ~0.09 ~&~ 2.845~\\
 ~0.1 ~&~ 3.150~\\
 ~0.2 ~&~ 6.268~\\
 ~0.3 ~&~ 9.364~\\
 ~0.4 ~&~ 12.428~\\
 ~0.5 ~&~ 15.461~\\
 ~1.0 ~&~ 30.171~\\
 ~1.5 ~&~ 43.238~\\
 \hline
\end{tabular}
\caption{Time period of oscillation ($T$) of kinetic and magnetic energy with $M_A$.}
\label{M_A_vs_omega}
\end{table}

%===========================

\subsubsection{Effect of Initial Wave-number ($k_0$)}

\begin{figure}
\includegraphics[scale=0.65]{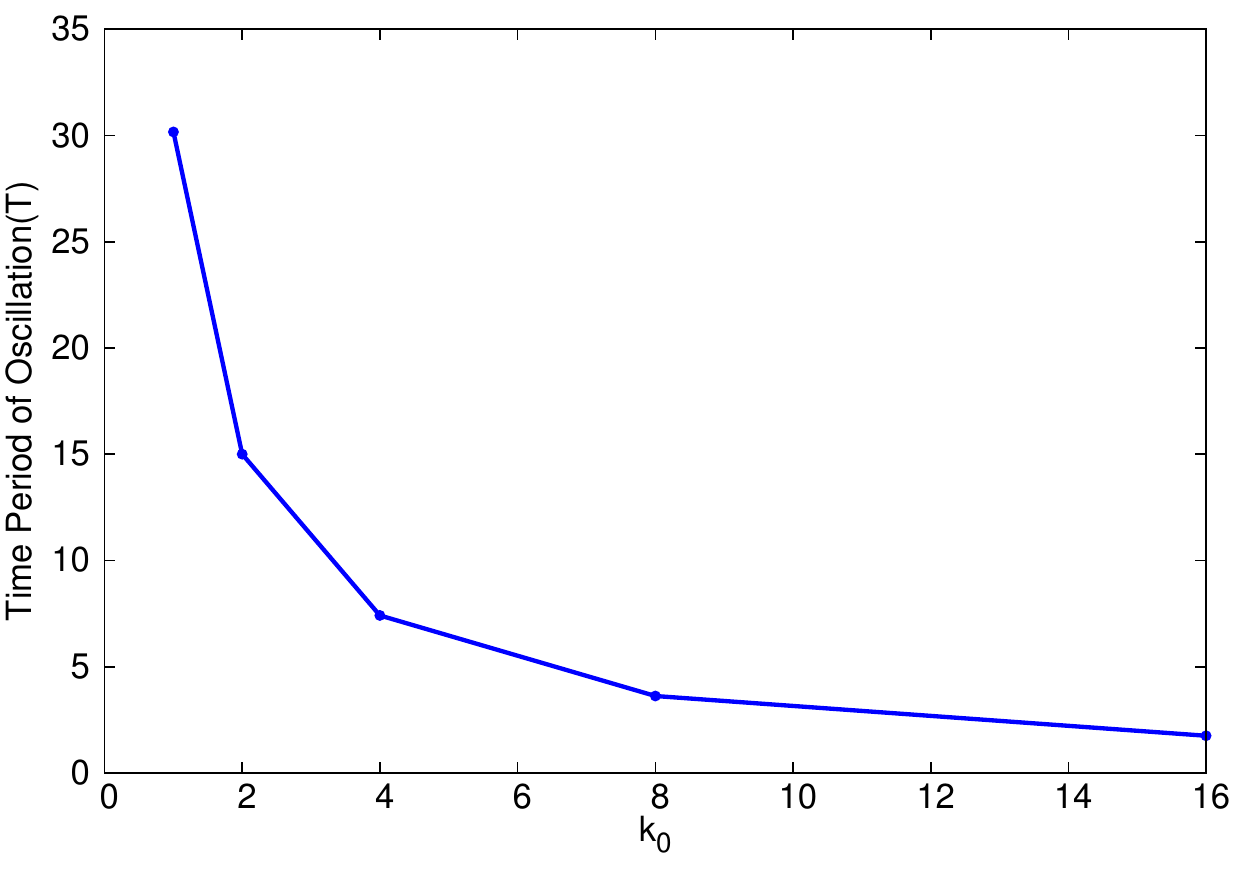}
\caption{Variation of time period of oscillation ($T$) of kinetic and magnetic energy with $k_0$. [see Table \ref{k_0_vs_omega}]}
\label{freq_k}
\end{figure}

We choose $k_0 = 1, 2, 4, 8, 16$ keeping $M_A = 1$. The frequency of oscillation increases with the increase of $k_0$ value keeping the phase velocity ($v_{phase} = \frac{\omega}{k}$) constant and equal to linear Alfv\'en velocity ($V_A$). This indicates that dispersion is absent for this oscillation. However the viscous ($\mu \nabla^2 \vec{u}$) as well as the resistive terms ($\eta \nabla^2 \vec{B}$) in Equation (\ref{velocity} and \ref{Bfield}) affects the dynamics more for higher wavenumber than for lower. It is because the second derivative in the viscous and the resistive terms become more prominent when the velocity and magnetic field variables contain a higher magnitude of $k_0$ [Table \ref{k_0_vs_omega}, Fig. \ref{freq_k}].\\

\begin{table}[h!]
\centering
\begin{tabular}{ |c|c| }
 \hline
 ~~~~$k_0$~~~~ & Time Period of Oscillation (T) \\
 \hline
 ~1 ~&~ 30.171~\\
 ~2 ~&~ 15.001~\\
 ~4 ~&~ 7.410~\\
 ~8 ~&~ 3.620~\\
 ~16 ~&~ 1.749~\\
 \hline
\end{tabular}
\caption{Time Period of oscillation ($T$) of kinetic and magnetic energy with mode number of excitation ($k_0$)}
\label{k_0_vs_omega}
\end{table}

\begin{figure}
\includegraphics[scale=0.65]{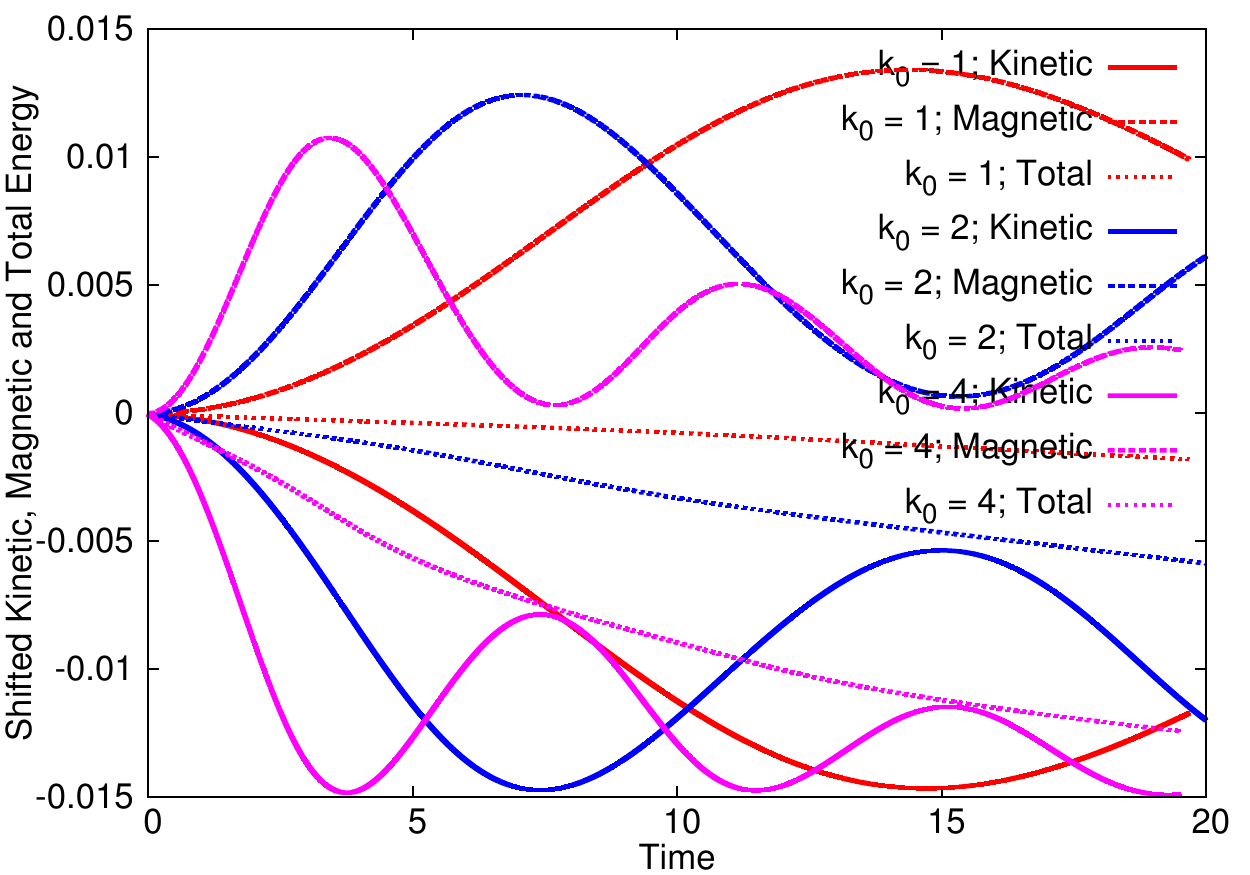}
\caption{Time evolution of shifted kinetic (solid), magnetic (dashed) and total (dotted) energy per grid for three dimensional Arnold-Beltrami-Childress flow with $N = 64^3$, $U_0 = 0.1$, $A = B = C = 1$, $k_0 = 1$ (red), $2$ (blue), $4$ (magenta), $M_s = 0.1$, $M_A = 1$ and $Re = Rm = 450$ in the absence of external forcing. Drag forces are found to affect the kinetic modes more than the magnetic modes as expected and explained in the text. The shift denotes the initial values of the kinetic and magnetic energies (at time $t = 0$) are subtracted from the time evolution data of the corresponding variables.}
\label{energy_unforced_kf_low}
\end{figure}

\begin{figure}
\includegraphics[scale=0.65]{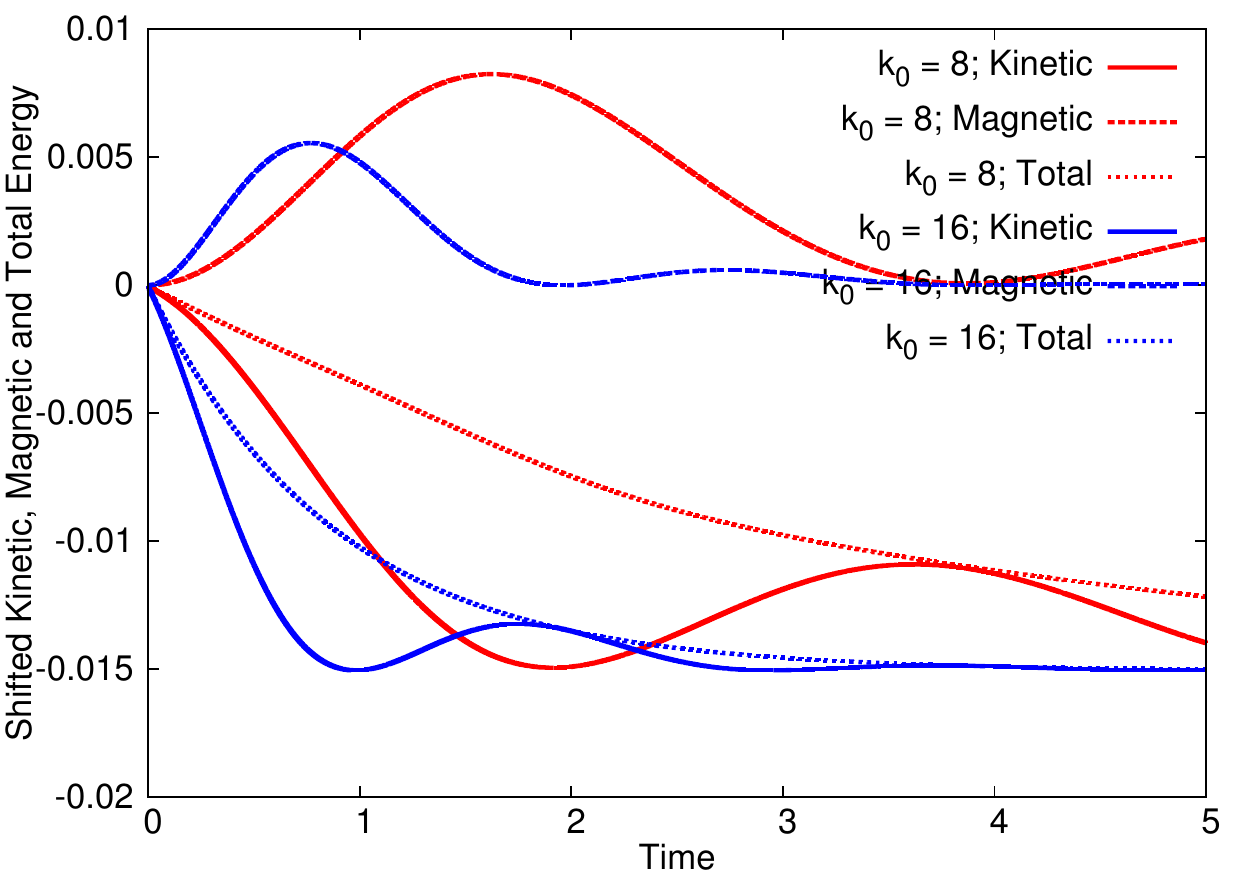}
\caption{Time evolution of shifted kinetic (solid), magnetic (dashed) and total (dotted) energy per grid for three dimensional Arnold-Beltrami-Childress flow with $N = 64^3$, $U_0 = 0.1$, $A = B = C = 1$, $k_0 = 8$ (red), $16$ (blue), $M_s = 0.1$, $M_A = 1$ and $Re = Rm = 450$ in the absence of external forcing. Drag forces are found to affect the kinetic modes more than the magnetic modes as expected and explained in the text. The shift denotes the initial values of the kinetic and magnetic energies (at time $t = 0$) are subtracted from the time evolution data of the corresponding variables.}
\label{energy_unforced_kf_high}
\end{figure}

%=========================================================================================

\section{Galerkin Representation in Two dimensions}
\label{sec:galerkin}
As already indicated, an exact solution for dispersionless nonlinear oscillation has been identified in the past \cite{alfven:1963}, here we present a Galerkin representation which, not surprisingly, further confirms our findings. Thus, we resort to a finite mode representation for the field variables for the two dimensional decaying case. We construct the stream function ($\vec{\Psi}$) defined as $\vec{u} = \vec{\nabla} \times \vec{\Psi}$ and magnetic vector potential ($\vec{A}$), defined as $\vec{B} = \vec{\nabla} \times \vec{A}$. Also, $\vec{\omega} = \vec{\nabla} \times \vec{u}$. In two dimensions, $\vec{\Psi} = (0,0,-\psi)$ and $\vec{u} = \hat{z} \times \vec{\nabla} \psi$.  Similarly, in two dimensions $\vec{A} = (0,0,-A)$ and $\vec{B} = \hat{z} \times \vec{\nabla} A$. Thus in two dimensions, $\vec{\omega} = \hat{z} ~ \nabla^2 \psi$ and $\vec{\nabla} \times \vec{B} = \hat{z} ~\nabla^2 A$. The normalised incompressible MHD equations are,
\begin{eqnarray*}
&& \frac{\partial \vec{u}}{\partial t} + (\vec{u} \cdot \vec{\nabla}) \vec{u} = (\vec{\nabla} \times \vec{B}) \times \vec{B}\\
&& \frac{\partial \vec{B}}{\partial t} = \vec{\nabla} \times (\vec{u} \times \vec{B})
\end{eqnarray*}
In two dimensions, the above equations can be expressed in $\psi$ \& $A$ variables as,
\begin{eqnarray}
&& \frac{\partial}{\partial t} \nabla^2 \psi  + (\hat{z} \times \vec{\nabla} \psi \cdot \vec{\nabla}) \nabla^2 \psi \nonumber \\
\label{velocity}&& ~~~~~~~~~~ = \hat{z} \cdot \vec{\nabla} \times \left[ \hat{z} ~ \nabla^2 A \times (\hat{z} \times \vec{\nabla} A) \right]\\
\label{bfield} && \frac{\partial A}{\partial t} + (\hat{z} \times \vec{\nabla} \psi \cdot \vec{\nabla}) A = 0
\end{eqnarray} 
Maintaining the incompressibility condition i.e. $\vec{\nabla} \cdot \vec{u} = 0 = \vec{\nabla} \cdot \vec{B}$, we represent the finite mode expansion as,

\begin{eqnarray}
&& \psi (x,y) = \psi_0 \sin kx + e^{iky} (\psi_1 + \psi_3 \cos kx) \nonumber \\
\label{psi} && ~~~~~~~~~~~~~~~~~~~~~~~~~ + e^{-iky} (\psi_1^* + \psi_3^* \cos kx)\\
&& A (x,y) = A_0 \sin kx + e^{iky} (A_1 + A_3 \cos kx) \nonumber \\
\label{A} && ~~~~~~~~~~~~~~~~~~~~~~~~~ + e^{-iky} (A_1^* + A_3^* \cos kx)
\end{eqnarray}
where $k$ is the wave vector corresponding to the largest scale length of the system. Substituting Equation \ref{psi} and \ref{A} into Equation \ref{velocity} and \ref{bfield} we obtain the time evolution of the galerkin truncated modes. The evolution of various modes of stream function and magnetic vector potential can thus be written as,

\begin{eqnarray}
&& \frac{d \psi_0}{dt} = i k^2 (\psi_3^* \psi_1 - \psi_3 \psi_1^*)\\
&& \frac{d\psi_1}{dt} = \frac{i}{2} k^2 (\psi_0 \psi_3 + A_0 A_3)\\
&& \frac{d\psi_3}{dt} = 0
\end{eqnarray}
\begin{eqnarray}
&& \frac{d A_0}{dt} = i k^2 (A_3 \psi_1^* - A_3^* \psi_1 + A_1 \psi_3^* - A_1^* \psi_3)\\
&& \frac{d A_1}{dt} = \frac{i}{2} k^2 (A_0 \psi_3 - A_3 \psi_0)\\
&& \frac{d A_3}{dt} = i k^2 (A_0 \psi_1 - A_1 \psi_0)
\end{eqnarray}

\begin{figure}
\includegraphics[scale=0.65]{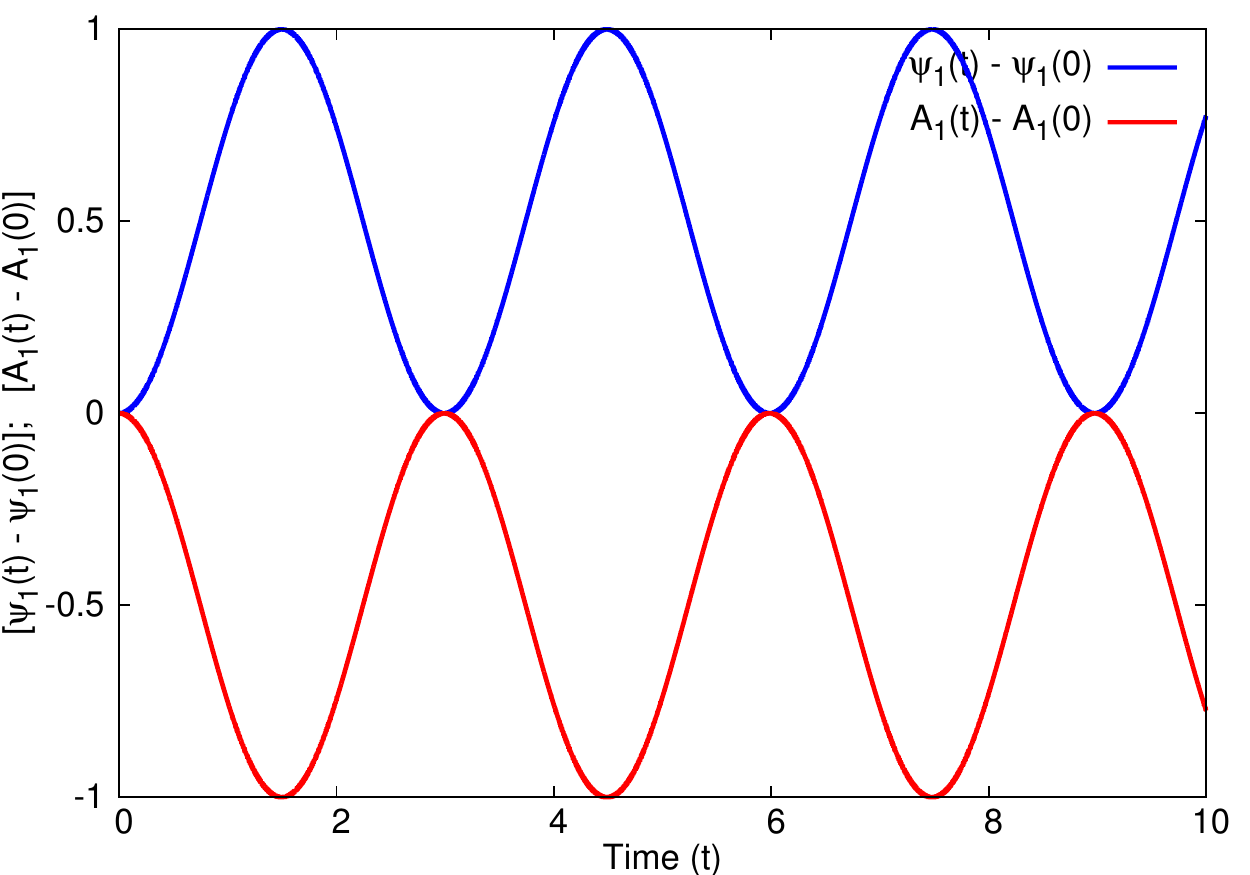}
\caption{Galerkin truncated two dimensional MHD equations representing similar nonlinear coherent oscillations observed in DNS results with $k = 1$.}
\label{galerkin}
\end{figure}

We start with a case similar to that of our initial condition used for the DNS code ($\psi_0 = 1$, $\psi_1 = 0 = \psi_3$, $A_0 = A_1 = A_3 = 1$ and $k = 1$) and time evolve the above set of ordinary differential equations using RK4 method. In Fig. \ref{galerkin} we show the time evolution of real part of $\psi_1$ and $A_1$. The variables ($\psi_1$ and $A_1$) are found to keep exchanging their values as the time evolves.

\begin{figure}
\includegraphics[scale=0.65]{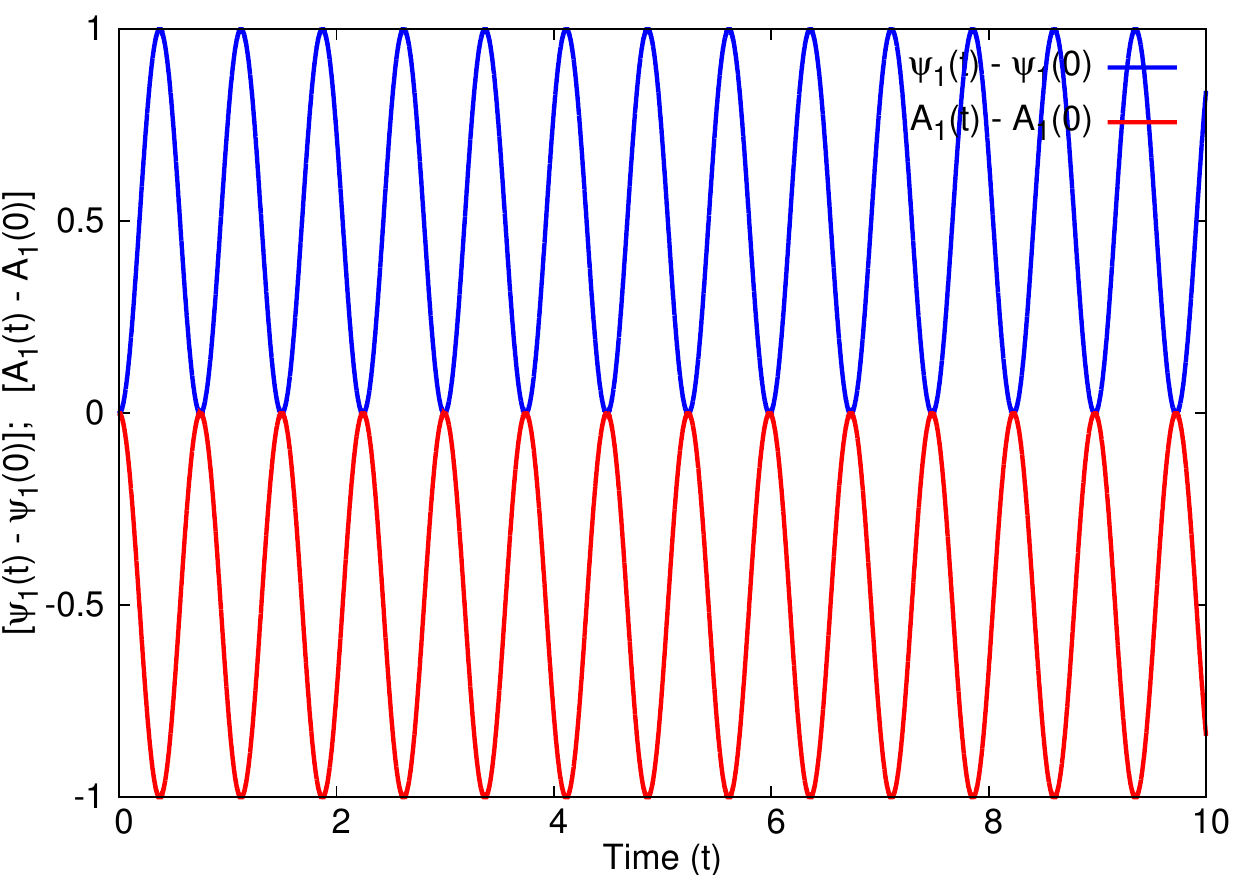}
\caption{Galerkin truncated two dimensional MHD equations representing similar nonlinear coherent oscillations observed in DNS results with $k = 2$.}
\label{galerkin_2}
\end{figure}

\begin{figure}
\includegraphics[scale=0.65]{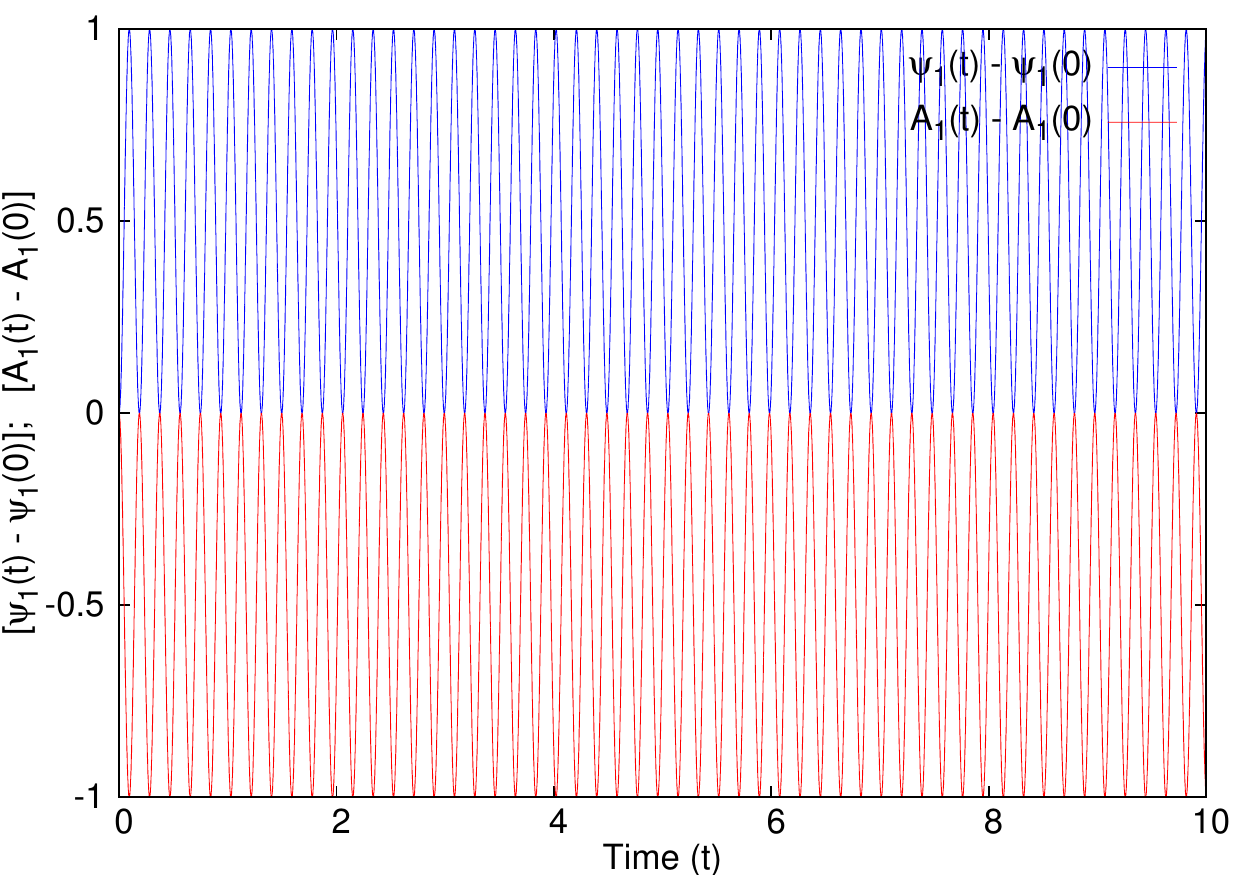}
\caption{Galerkin truncated two dimensional MHD equations representing similar nonlinear coherent oscillations observed in DNS results with $k = 4$.}
\label{galerkin_4}
\end{figure}

We also find the increment of frequency with the increase of $k_0$ in two dimensional system similar to that in three dimensional system [Fig \ref{galerkin_2},\ref{galerkin_4}.

%=========================================================================================

\section{Summary and Future Work}
\label{sec:summary}
In this paper we report numerical evidence of the existence of nonlinear non-dispersive Alfv\'en waves in a model system of  single fluid magnetohydrodynamic equations. These waves provide a coherent mechanism of large-amplitude energy exchange between the kinetic and magnetic energy components of the system and sustain themselves  over very long periods in both two and three dimensions. The time-period of oscillation of the energy exchange is found to vary linearly with Alfven Mach number ($M_A$) as expected for linear Alfv\'en waves. Also the frequency of oscillation is seen to increase linearly with the initial wave-number ($k_0$) excited within the flow keeping the phase velocity of the wave constant. This suggests that the waves excited are dispersionless. Furthermore, starting from a non-sinusoidal velocity profile in the presence of ambient uniform magnetic field, it is observed that the nonlinear Alfven waves do not get excited into the medium. Also, by reversing the nature of initial condition between flow field and magnetic field, it is found that the oscillations disappear.\\

The three dimensional calculations are performed over $64^3$ grid resolution. A better spatial resolution is expected to be capable of simulating higher fluid and magnetic Reynold's number runs. Thus the periodicity of energy exchange in three dimensional flows could be observed for larger number of cycles for high resolution runs.

%=========================================================================================

\section{Acknowledgement}
The development as well as benchmarking of MHD2D and MHD3D has been done at Udbhav and Uday clusters at IPR. R.M. thanks Samriddhi Sankar Ray at International Center for Theoretical Sciences, India for his initial help regarding pseudo spectral simulation and Sayak Bose, Columbia University for several helpful discussions.   The support from ICTS program: ICTS/Prog-dcs/2016/05 is also acknowledged. R.M. acknowledges the financial assistance from Department of Atomic Energy (DAE) India, under the PhD Fellowship scheme. A.S. is thankful to the Indian National Science Academy (INSA) for their support under the INSA Senior Scientist Fellowship scheme.

%=========================================================================================

%\nocite{*}

\bibliography{biblio}

\end{document}